\documentclass{sig-alternate}

\usepackage{graphicx}
\usepackage[hyphens]{url}
\usepackage{url} 
\usepackage{hyperref}
\usepackage[hyphenbreaks]{breakurl}
\usepackage{rotating}

\usepackage{times}
\usepackage[scaled]{helvet}
\makeatletter
\def\@copyrightspace{\relax}	
\makeatother
\begin{document}
\title{A Case of Collusion: A Study of the Interface Between Ad Libraries and their Apps}
\numberofauthors{2}
\author{
\alignauthor
Theodore Book\\
\affaddr{Rice University}
\email{\textsf{tbook@rice.edu}}
\alignauthor
Dan S. Wallach\\
\affaddr{Rice University}
\email{\textsf{dwallach@cs.rice.edu}}
}

\maketitle

\abstract{
A growing concern with advertisement libraries on Android is their ability to exfiltrate personal information from their host applications. While previous work has looked at the libraries' abilities to extract private information from the system, advertising libraries also include APIs through which a host application can deliberately leak private information about the user. This study considers a corpus of 114,000 apps.  We reconstruct the APIs for 103 ad libraries used in the corpus, and study how the privacy leaking APIs from the top 20 ad libraries are used by the 64,000 applications in which they are included. Notably, we have found that app popularity correlates with privacy leakage; the marginal increase in advertising revenue, multiplied over a larger user base, seems to incentivize these app vendors to violate their users' privacy.
}
\section{Introduction}
The Android operating system is one of the primary mobile device platforms worldwide, accounting for 70\% of smartphone shipments~\cite{strategy-android-market-share}.  It also supports a vibrant advertising industry, with dozens of advertising agencies providing ad libraries installed in hundreds of thousands of free applications.  Indeed, for every paid app downloaded on Google Play, users download 82 free apps, mostly ad supported~\cite{DistimoPaidVsFree}.  Privacy threats on Android are similar to those on iOS, where apps exhibit similar privacy leakages~\cite{sw-ios-android}.  The diversity of these libraries and the possibility of studying their deployment in real world applications makes Android an ideal platform to study the impact of advertising libraries on user privacy and mobile device security.

A great deal of recent research has focused on the relationship between advertising libraries and the Android operating system, with a particular focus on the use of Android API calls protected by permissions~\cite{book2013longitudinal, stevensinvestigating, grace2012unsafe}.  However, very little attention has been paid to the other interface of ad libraries: the API that they use to interact with their host application.  This interface represents a significant privacy concern, because host applications have access to confidential user data that extends beyond the information that they request from the operating system.  Indeed, applications may have access to a great deal of confidential data, obtained through system calls, direct user input, and social network APIs, as well as data shared by other applications, on the device or in the cloud.

Generally speaking, the application developer's consent is necessary for an ad library to access this trove of information.  Barring creative run-time inspection by the library of the host application's data structures, it is up to the developer to pass data to the ad library through the ad library API.  However, developers have every incentive to hand personal data to advertising agencies, as it has the potential to increase advertising revenue for their applications.  In this study, we set out to understand these interfaces, and the extent to which developers make use of them.

\section{Methodology}

\begin{figure}
\centering
\includegraphics[width=\columnwidth, trim = 0in 0.3in 0in 0in]{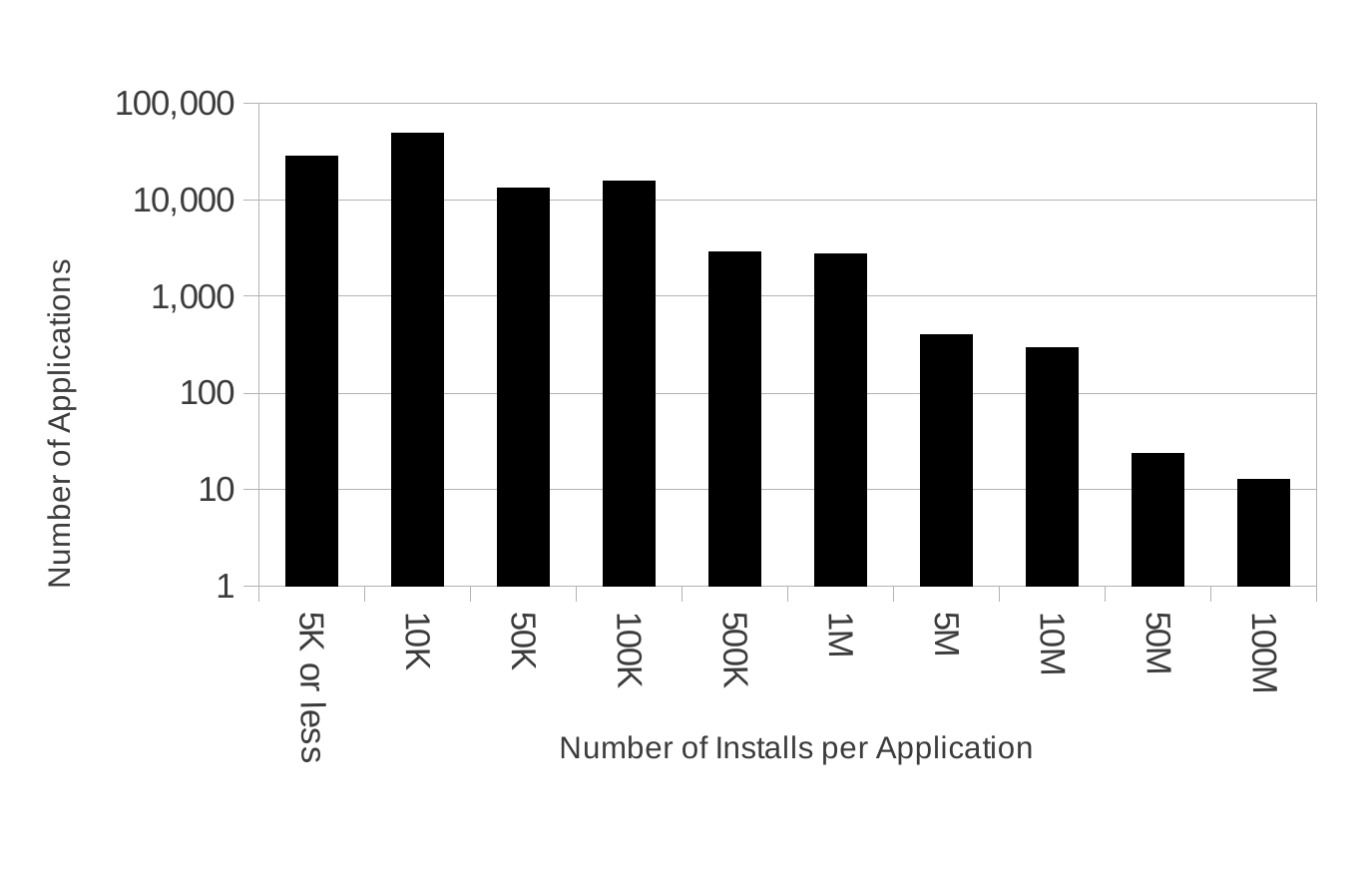}
\caption{Sample broken down by install count}
\label{apps-by-installs}
\end{figure}

We began with a collection of 114,000 apps downloaded from Google Play in early 2013, the same that was used in our earlier work on ad library permission usage~\cite{book2013longitudinal}.  This collection consists exclusively of free applications, with a focus on the most popular applications on Google Play.  Figure~\ref{apps-by-installs} shows the distribution of the sample according to the number of installs per application.  Because the sample focuses on the most popular applications, we believe that it includes a large portion of the applications likely to be found on an average end-user's device.  We reconstructed the APIs for these libraries, identified privacy-related API calls for the top 20, and studied how those calls were used in applications.

\subsection{API Extraction}
Within this collection, we manually identified the package names for 103 individual ad and analytics libraries.  We then disassembled all of our applications using Gabor Paller's Dedexer.\footnote{http://dedexer.sourceforge.net}  Using the package names, we parsed through our applications to identify all API calls made from any application to one of its ad libraries.

By assembling the set of all observed API calls, we were able to reconstruct the API for each ad library, generating a complete record of all API calls that were actually used within our dataset.  We were also able to track the frequency with which each API call is used.  In extracting data from the apps, we were able to retrieve not only method names, but also parameter and return types, giving us a profile of the interface between the application and its libraries.  We did not consider different versions of the libraries, choosing to differentiate libraries at the level of package name.  This means that if a certain API call was only included in some versions of a library, we still included it in our API reference.  When we measure frequency of use, we measure it across all versions of a library, even if a given call might not be included in some versions. 

In theory, there are other ways that an app can communicate with its ad library aside from API calls---direct manipulation of class variables, shared memory, calls to the app from the library, etc.---but the need to interact with a variety of developers effectively prohibits the use of more indirect techniques.  Nonetheless, our detection of interactions based on API calls does not capture any such interactions, and thus may only study a subset of the information flowing from applications to their ad libraries.

On the Android platform there is one additional significant interface between the application and the ad library---the Android XML layout files included in the application.  These layout files allow the developer to instruct the library to display an ad, and to pass various parameters.  However, because the parameters are static and fixed at build time, their ability to pass significant personal information is limited.  Nonetheless, to the extent that the app developer is able to predict the demographic of his users, the XML layout file does present a potential point of information leakage.  Indeed, it seems that some ad library designers have sought to exploit this interface: at least one ad library, Jumptap, allows developers to embed demographic information, including age, gender, household income, and postal code, in the static XML files for the application~\cite{jumptapAPI}.

\noindent {\bf Obfuscation.}
One advertising library, AirPush, recently adopted a scheme to obfuscate package names.  Each developer receives a binary with a unique package name when they download the library.  However, as the library code remains the same, we were able to determine some characteristic features that allowed us to recognize instances of the AirPush library despite the obfuscated name.

The use of obfuscation by application developers was more difficult to process.  In some cases, ad libraries were re-written by obfuscation and optimization software, changing the method names.  In these cases, we were unable to detect privacy related API calls.  An analysis of calls to AdMob indicated that approximately 5\% of API calls were obfuscated in this way.  Given this, our findings represent a lower bound on the number of applications leaking privacy related information through ad library API calls.

\subsection{API Analysis}
Once we had reconstructed the API for each ad library, we set out to identify privacy related API calls.  We did this by a manual inspection of each API call, using the method name and parameters to form an initial evaluation of whether the API call constituted a privacy risk.  In cases of uncertainty, we turned to official API references and sample code available on the Internet in order to form a better assessment.  While it would be trivial for a devious library developer to choose method names that would confound this sort of analysis, of necessity the API is designed to be accessible to application developers, and so such a choice would make it difficult for developers to use the API, causing it to fail in its primary purpose.  Anecdotally, looking at the API calls that we identified, there was no indication that library designers sought to present method names that hid their true function.

We conducted this analysis on the top 20 libraries, which together account for 84\% of all installs.  64,000 apps within our corpus contained at least one of these libraries.  By excluding the smaller libraries, we necessarily failed to detect some API calls that present the potential for privacy leaks, as well as some of the applications that made use of those APIs.

In order to understand how these API calls are used by applications, we then counted the privacy related ad library API calls made by each app, recording the number of apps making use of each type of call.  This enabled us to determine the percentage of apps passing personal information directly to ad libraries, and so to quantify the scale of the privacy concern represented by these libraries. Our analysis has its limits of course:

\noindent {\bf Key/value maps.} Some ad libraries allow passing general key/value pairings (e.g., Java ``map'' objects), allowing a variety of different privacy-relevant items to be passed at once. We did not implement the static analysis and analysis over Dalvik bytecode that would be necessary to track these values. All we see is that the general-purpose call is made, not what keys are included.  In our results, we report use of these APIs in a separate category.

\noindent {\bf Ad mediation libraries.} Another complication arises from ad mediation libraries, which switch among different advertisers. Since we're interested in leaks from applications to ads, and not from mediation libraries to other ads, we excluded calls from one ad library to another. Note that this methodology also excludes information paths where one ad library retrieves information from the underlying Android system (for example, by querying the device's location) and then passes it to another ad library.

\section{Results}

For developers, the multiplicity of ad libraries on the Android platform provides options, allowing them to select libraries based on privacy concerns as well as functionality and revenue potential.  However, for users, who do not choose the ad libraries in their apps, this represents expanded vulnerability, as the presence of any library that exfiltrates personal data means that the data is potentially compromised.  We show that many libraries have the potential to do exactly that.

\subsection{Library market share}

\begin{figure}
\centering
\includegraphics[width=\columnwidth, trim = 0in 0.1in 0in 0in]{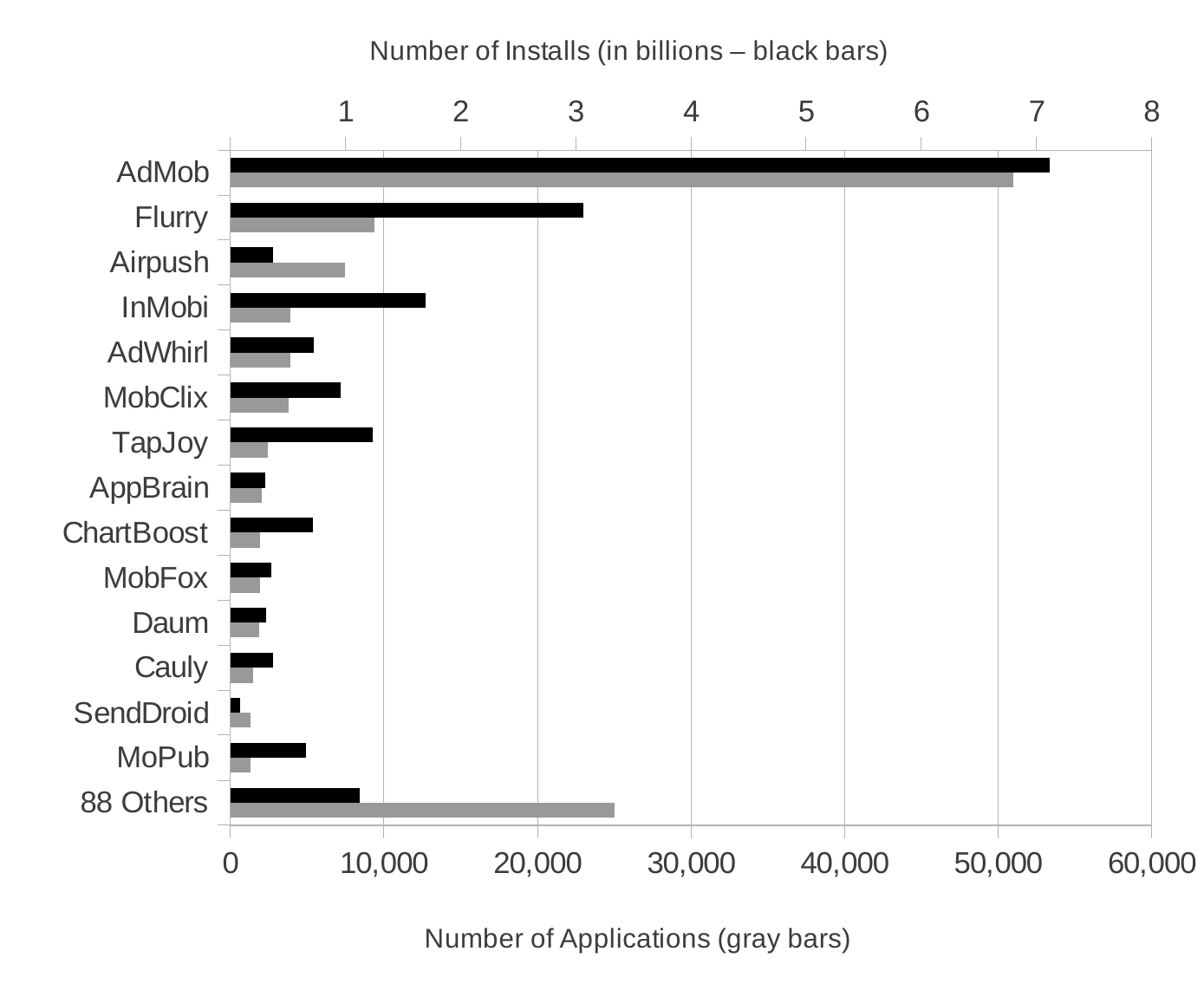}
\caption{Ad Libraries by Number of Apps}
\label{libs-by-appcount}
\end{figure}

In order to determine which of our 103 libraries to focus on for our analysis, we needed to understand the relative market share of each library.  We did this by simply enumerating the number of installs for each library and comparing it to the total number of apps in our dataset.  Results are displayed in figure~\ref{libs-by-appcount}.  Both package names for AdMob (\textsf{\small{com.google.ads}} and \textsf{\small{com.admob}}) are combined into the AdMob listing.  Numbers for AirPush include both obfuscated and non-obfuscated package names.  As can be seen, AdMob has a commanding market share, but other libraries represent around 65\% of installs.  It is likely that we failed to identify some advertising libraries, but it seems probable that any such libraries are not among the most common.  Nonetheless, if they were included, the ``Other'' category would probably larger than indicated in Figure~\ref{libs-by-appcount}.

It is difficult to estimate the number of libraries that are likely to be found on a ``typical'' device.  Our survey of 114,000 apps identified 119,000 ad library instances, or an average of about one per app.  However, some apps contain many libraries, and some contain none.  The number of libraries on a user's device is probably highly dependent not only on how many apps he chooses to install, but also on which apps he selects.  Anecdotally, certain families of apps, such as live wallpapers and pornographic applications, seem to tend to contain large numbers of ad libraries.  Others, such as applications designed to interact with a business with which the device user has an existing relationship, such as banking or airline applications, tend to contain no ad libraries at all.  

\subsection{Ad Library APIs}

While many APIs have publicly available documentation, many others restrict access to their API documentation to registered developers.  Additionally, our method of extracting calls actually used by applications allows us to understand the ``working API''---those calls, documented or not, that are actually used in applications, together with their popularity.  While app developers' use of obfuscation software that modifies method names makes it somewhat difficult to count accurately the number of API calls in a given library, it is worth noting that many have a very simple API, with 17 of 103 libraries having 10 or fewer API calls.  Promotional literature for ad libraries often stresses the ease of incorporating the library into an application, which may account, in part, for the small APIs.

The most common API calls for nearly all libraries are, as might be expected, calls related to laying out and requesting advertising content.  To take AdMob as an example, the constructor for the AdRequest object is the most frequently called method.  After this, the loadAd method for an AdView follows closely, together with the constructor for an AdView.  These are followed by various layout and lifecycle methods.  Other libraries have a similar distribution of calls.

\subsection{Privacy related API calls}
\begin{table}
\centering
\begin{tabular}{ l | c | c}
Classification & Percent of Apps & Percent of Installs\\
\hline
Arbitrary Data & 3.06\% & 9.13\%\\
Keywords & 2.50\% & 5.87\%\\
Gender & 2.03\% & 3.06\%\\
Location & 1.64\% & 3.38\%\\
Age & 1.50\% & 2.66\%\\
Multiple Factors & 0.50\% & 1.99\%\\
Postal Code & 0.42\% & 0.49\%\\
Enable Location & 0.34\% & 0.32\%\\
Income & 0.12\% & 0.07\%\\
Interests & 0.01\% & 0.01\%\\
Area Code & 0.01\% & 0.01\%\\
Country & 0.01\% & 0.12\%\\
Education & 0.01\% & 0.01\%\\
Ethnicity & 0.00\% & 0.00\%\\
Name & 0.00\% & 0.00\%\\
E-Mail & 0.00\% & 0.00\%\\

\end{tabular}
\caption{Percentage of apps making a call to a top 20 library}
\label{call-use-all}
\end{table}

This study is not concerned with the most common calls, but those that represent potential privacy leaks.  Of the top 20 libraries, 11 include calls in their APIs that have the potential to leak user data.  The sort of data leaked included everything from location to household income.  For a list of all factors identified, see Table~\ref{call-use-all}.  

Most of these categories are self-explanatory, but a few deserve additional detail.  
The category labeled ``Arbitrary Data'' refers to calls that allow the developer to send arbitrary data to himself.  These calls are common in libraries with analytics functions, and are presumably intended to be used to transfer information regarding the usage and performance of the application.  They could, however, transfer any sort of data, and so represent a potential privacy leak.  However, because a developer can send information to an arbitrary server without the need for an advertising library, their presence in advertising and analytics libraries does not represent an additional threat.

The category ``Age'' encompasses both API calls that give the user's age in years and API calls that transmit the user's exact birth date.  While both have similar value in targeting advertisements, a user's exact birth date provides a much greater risk of de-anonymization.

The category labeled ``Multiple Factors,'' on the other hand, represents API calls that permit the sending of various pieces of demographic data to the ad agency.  These calls generally accept a key/value store, such as a Java map, which can contain a variety of factors, generally similar to the ones that can be passed through individual calls.  While we did not analyze which factors are passed through these calls by which applications, the relative infrequency of their use makes such counts unnecessary for understanding the overall data flow through the ad library APIs.

The category ``Enable Location'' is also slightly different from the other calls.  Rather than passing information to the library, it authorizes the library to make (or not make) system calls to collect location data directly.  Thus, the presence of this call controls a location function that is present in the library.

As can be seen, the privacy related data that is leaked through ad library API calls is data that might be useful in targeting advertising.  A few categories, such as postal code and area code, seem to be carry-overs from the world of print and phone marketing, and perhaps point to an advertising model where information associated with a user, outside of the phone through some other business relationship, is being send through the phone to the advertising agency.

Without access to internal data from the ad agencies, it is difficult to know to what extent this information is used to uniquely identify users or to link their identities with external databases.  However, previous research has shown that as of the year 2000, 63\% of the US population could be uniquely identified by gender, zip code, and date of birth, without reference to other factors~\cite{golle2006revisiting}.   Given that many of these libraries collect other identifying factors beyond these basic demographic indicators, it would appear that some are collecting a sufficient amount of data to de-anonymize users.

When one remembers that libraries can also collect information directly from the underlying Android system, and that ad agencies can use unique device identifiers to track users from one application to another, it becomes clear that the amount of data collected on an individual user could be significant.  When combined, this could again be used to de-anonymize users.

It is also worth noting that most of these calls transfer information that the libraries could not themselves obtain via privileged Android calls.  Thus, they extend the dataset that an ad agency is able to build around a user by bringing in information from other sources.  While this may be information directly entered by the user into the application, it can also be information from social networking sites or other cloud data sources that the user authorizes the application to access.  

\subsection{Number of applications making privacy related API calls}

Table~\ref{call-use-all} shows the percentage of the total universe of apps that make a call in a given category to one or more of the top 20 libraries.  It can be seen that certain API calls are called much more frequently than others.  Aside from the ``Arbitrary Data'' category, whose seriousness depends on the way that individual developers choose to use it, the most common category is keywords, which may relate to user activity inside an app, but is not fundamentally different from the sort of personal data habitually collected in web advertising.  The next most common category is gender, which, together with age, provides demographic information without being, by itself, personally identifying  (It should be noted, however, that the age category includes API calls that transmit birth dates, which go a long way towards uniquely identifying an individual.)  Location information is also frequently passed to ad libraries, although many libraries also collect this information directly from the Android operating system~\cite{book2013longitudinal}.

The only other pieces of data that are collected with any frequency are postal code and income.  While the value of income for ad targeting is obvious, it is interesting to speculate on the value of post code information in situations where the user's current location is known.  It undoubtedly allows the ad agency to link the user to a great deal of post code specific demographic information, but it also provides a significant vector which, when combined with others, could be used to de-anonymize users.

There are several pieces of data which are almost never shared with ad libraries, some of which, such as name and e-mail, are nearly sufficient by themselves to de-anonymize users.  The scarcity of the use of these API calls may be attributed to their presence in only one less-popular library, as well as the rarity of their invocation even in apps that include that library.  Perhaps this represents a sense of ethics on the part of library and app developers, or perhaps it simply represents a pragmatic choice regarding the usefulness of that information combined with the costs of collecting it.

\subsection{Popularity of applications making privacy related API calls}

\begin{figure}
\centering
\includegraphics[width=\columnwidth, trim = 0in 0.3in 0in 0.2in]{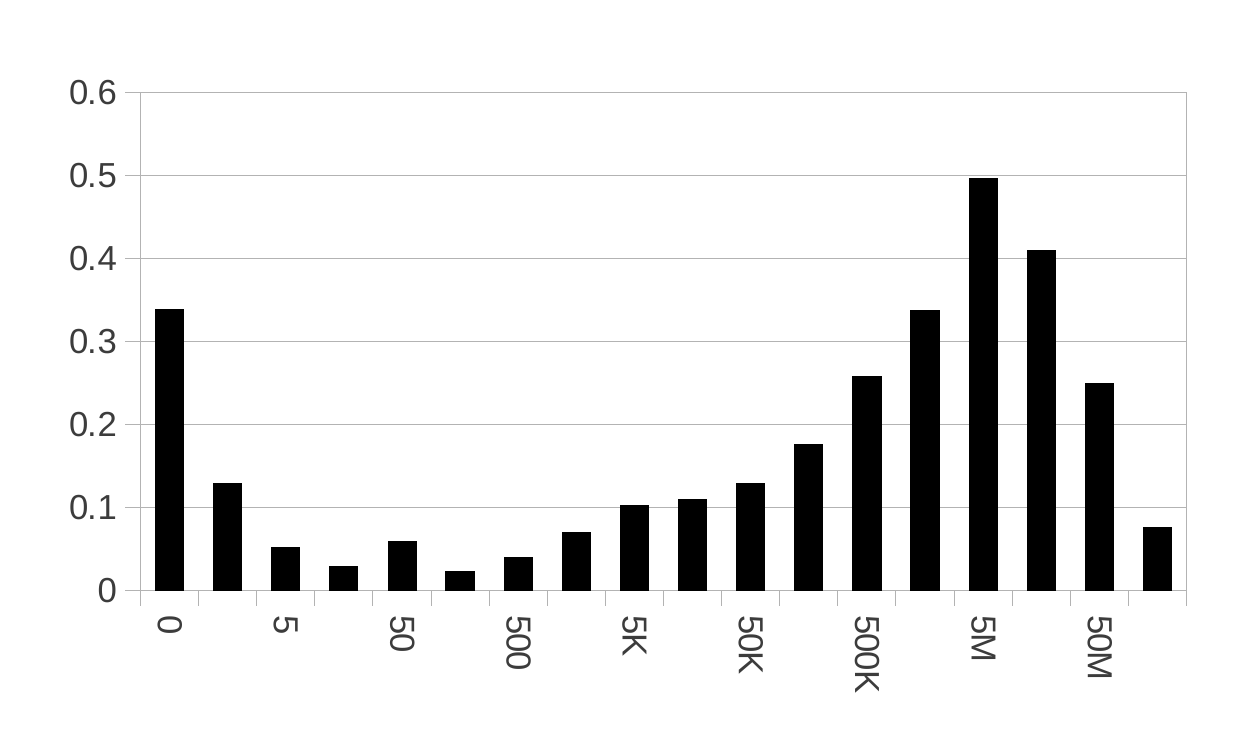}
\caption{Average number of privacy related API calls per app by number of installs}
\label{apps-by-popularity}
\end{figure}

An additional question that comes to mind when considering the use of privacy-related API calls concerns the popularity of applications using these calls.  Are these rarely-installed rogue applications, or do the represent the ``mainstream'' of Android apps---those with a large number of installs?

Figure~\ref{apps-by-popularity} shows the number of unique privacy related calls per app.  The apps are divided into the usage categories given by Google Play, with the lower bounds of the category given for each app.  Thus, we assume 5,000 installs for an app that Google lists as having 5,000 to 10,000 installs.  If a single app makes multiple calls of the same type to the same library, each set of calls is only counted once.  However, calls sending different sorts of information to the same library, or calls sending the same sort of information to different libraries are counted separately.

As can be seen from Figure~\ref{apps-by-popularity}, there is a bimodal distribution, where the least and most popular apps are leaking the most privacy-relevant data to ad libraries. These leaks peak at nearly 0.5 calls per app in apps with 5 million to 10 million installs. While our sample contains only a small number of apps with fewer than 5,000 installs, making the data for the lower end of the spectrum less accurate, it contains most of the free applications with at least 10,000 installs, providing very accurate data for those categories.

It is reasonable to assume that the applications with the largest number of installs are also the applications which have received the largest amount of development resources.  They are also the applications whose developers have the most to gain from marginal increases in ad revenue for user.  Given these facts, it is perhaps not surprising that these applications are the ones which are most likely to contain privacy related API calls.  However, it also indicates that the average app install is more likely to contain these calls than might be otherwise expected, as the more popular apps are proportionally more likely to be installed on any given device.

\subsection{Correlation with permission usage}

\begin{table}
\centering
\begin{tabular}{ l | c c }
Library & Permissions & Number of API calls \\
\hline
 AdMob & 5 & 5 \\
 Flurry & 5 & 4 \\
 Airpush & 9 & 2 \\
 InMobi & 7 & 12 \\
 AdWhirl & 4 & 3 \\
 MobClix & 15 & 0 \\
 TapJoy & 5 & 0 \\
 AppBrain & 2 & 0 \\
 ChartBoost & 4 & 0 \\
 MobFox & 6 & 0 \\
 Daum & 11 & 3 \\
 Cauly & 7 & 1 \\
 SendDroid & 8 & 0 \\
 MoPub & 5 & 3 \\
 Google Analytics & 2 & 1 \\
 JumpTap & 6 & 6 \\
 AppLovin & 9 & 8 \\
 Mediba & 4 & 0 \\
 Inneractive & 6 & 0 \\
 GreyStripe & 4 & 0 \\
\end{tabular}
\caption{Permission Usage and Privacy Related API calls}
\label{perm-use}
\end{table}

Ad libraries can obtain sensitive information from system calls as well as from their own API calls. 
Our previous research documented the number of system permissions that ad libraries were able to exploit~\cite{book2013longitudinal}.
  We desired to know whether there was any correlation between these two forms of intrusive behavior.  Is it the case that libraries that use large numbers of permissions also seek to gather personal information through developer accessible API calls, or does the use of one sort of information replace the need for the other?

Table~\ref{perm-use} compares the number of API calls and the maximal number of permissions used by any version of a given library.  We found that there was a very low correlation coefficient (0.14) between the two vectors.  Some libraries made abundant use of both interfaces, while others primarily made use of one or the other.  The decision to make use of one interface or the other appears to have been based on independent decisions by the developers.

\subsection{Library Comparison}
\begin{table*}
\begin{minipage}{\textwidth}
\centering
\begin{tabular}{ l | c c c c c c c c c c c c c c c c c c c c c}
&
\begin{sideways}AdMob\end{sideways} &
\begin{sideways}Flurry\end{sideways} &
\begin{sideways}AirPush\end{sideways} &
\begin{sideways}InMobi\end{sideways} &
\begin{sideways}AdWhirl\end{sideways} &
\begin{sideways}MobClix\end{sideways} &
\begin{sideways}TapJoy\end{sideways} &
\begin{sideways}AppBrain\end{sideways} &
\begin{sideways}ChartBoost\end{sideways} &
\begin{sideways}Daum\end{sideways} &
\begin{sideways}Cauly\end{sideways} &
\begin{sideways}SendDroid\end{sideways} &
\begin{sideways}MoPub\end{sideways} &
\begin{sideways}Google Analytics\end{sideways} &
\begin{sideways}SendDroid\end{sideways} &
\begin{sideways}JumpTap\end{sideways} &
\begin{sideways}AppLovin\end{sideways} &
\begin{sideways}Mediba\end{sideways} &
\begin{sideways}Inneractive\end{sideways} &
\begin{sideways}GreyStripe\end{sideways} &
 \\
\hline
  Name & & & & & & & & & & & & & & & & & 0.1\\
  Location  & 2.7 & 4.8 & 0.0 & 1.3 & & & & & & & & & 4.8 & & & & 0.2\\
  Gender   & 2.5 & 3.4 & & 2.4 & 15.6 & & & & & 0.0 & & & & & & 13.8 & 0.5\\
  Age  & 1.5 & 3.4 & & 1.8 & 13.7 & & & & & 0.1 & & & & & & 13.8 & 0.5\\
  Education & & & & 0.2 & & & & & & & & & & & & &\\
  Ethnicity & & & & 0.1 & & & & & & & & & & & & &\\
  Income & & & & 0.1 & & & & & & & & & & & & 12.8 &\\
  Postal Code & 0.0\footnote{The older com.admob package included this feature.  The newer com.google.ads does not.} & & & 0.4 & 9.1 & & & & & & & & & & & 9.9 &\\
  Area Code & & & & 0.2 & & & & & & & & & & & & &\\
  Country & & & & & & & & & & & & & & & & 0.6 & 0.3\\
  Interests & & & & 0.3 & & & & & & & & & & & & & 0.1\\
  E-Mail & & & & & & & & & & & & & & & & & 0.1\\
  Keywords & 3.2 & & & 0.6 & 27.7 & & & & & & & & 12.7 & & & & 0.3\\
  Arbitrary Data\footnote{To the app developer} & & 31.1 & & & & & & & & & & & & 47.5 & & &\\
  Multiple Factors & & & 0.0 & 1.7 & & & & & & & 34.9 & & & & & &\\
  Enable Location & & & & 0.5 & & & & & & 1.4 & & & 0.5 & & & 14.3 &\\
\end{tabular}
\caption{Top 20 Libraries: Percentage of apps making privacy related API calls}
\label{call-use}
\end{minipage}
\end{table*}

Looking at our data in greater detail, we can examine the varying behaviors of the different libraries.  Table~\ref{call-use} shows the percentage of apps using a given library that make use of the various API calls.

It can immediately be seen which libraries provide API calls allowing developers to share privacy related data.  Some offer a much broader interface than others.  Of course, the cooperation of the developer is necessary for the ad agency to receive this information.  Our results show that most developers either do not have access to this information or choose not to share it.

It is worth noting that nine out of the top twenty libraries---that is to say, nearly half---do not appear to provide any API for developers to pass personal data to the library.  While these libraries may present other privacy or security related concerns, they present no risks in this particular area.

We can see that privacy-related API calls are invoked more frequently for some libraries.  While we do not have any data that would directly explain this phenomenon, some agencies may do more to encourage developers to provide personal information.  There might also be some correlation between developers who use certain libraries and those who desire to leak personal data.

It is important to note that we are only counting calls made directly from the application, and not calls that are made from one library to another.  This may account for the much higher rate that these API calls are made to AdWhirl.  AdWhirl is an ad mediation library, which allows a developer to incorporate various ad libraries in an application, and display ads from one of the networks based on factors such as ad availability and value.  In this way, information passed to AdWhirl may be passed, in turn, to the other libraries called by AdWhirl.

\section{Discussion}

Although we have determined that the percentage of applications which pass personal information to ad libraries is relatively low, the impact of this interface may be greater than one might first think.  Once the information is transmitted to the ad agency, it does not disappear, but is presumably logged in a database.  Because ad libraries often transmit a unique device ID, this information can then be correlated to information transmitted by other applications using the same ad library, and then be used in targeting ads to the user.  To the extent that the user's identity may be revealed by the ad libraries, this data can then be correlated to other demographic databases, not only allowing data from those databases to be used for targeting ads on the mobile device, but also allowing those databases to be updated with data obtained from the mobile libraries (for example, the location information that may be obtained through the library, the system, or the user's IP address).

One factor that is not addressed in this study, and that must be left to future work, is the question of the accuracy of the data provided by app developers to ad library vendors.  They might infer certain demographic characteristics of the user from the nature of the application, itself---for example, a children's game could have a hard-coded age value that it provides to the advertising library.  There might also be instances where applications entirely falsify keywords or demographic information in the hopes of obtaining greater revenue from the ad agencies.  We have not, however, investigated this possibility.

Much existing research into mitigating the privacy risk of Android ad libraries has focused on restricting the ability of ad libraries to make use of application permissions~\cite{pearce2012addroid, shekhar2012adsplit}.  It is worth noting that these techniques do not address the privacy risks posed by data passed through the library API.  On the other hand, approaches that use static analysis to identify privacy and security concerns can easily detect this behavior, as do methods that look for malicious behavior on the level of the application, without considering the distinction between application and library~\cite{rosen2013appprofiler, zhou2012hey}.  These methods can be automated and applied on a large scale~\cite{gibler2012androidleaks}.  

\noindent {\bf Future work.}  On the technical front, we expect more opportunities to study app behavior in the wild, either through static analysis, dynamic emulation, or on instrumented handsets. Likewise, our work here has focused on apps from Google Play, which must pass Google's deliberately vague standards. Apps from other platforms might behave worse. Also of interest are new technical means for users to control the behavior of their apps, such as CyanogenMod's experimental new Privacy Guard Manager\footnote{\url{https://plus.google.com/+CyanogenMod/posts/86LLXrDpVWY}}, which allows users to restrict apps access to location, contacts, and so forth, regardless of those apps' required permissions.

On the policy front, what level of privacy (required notifications, explicit consent) should users expect on their mobile devices?  What standards should govern ad agencies and application developers seeking to collect personal data and  who should administer them?  Additionally, awareness of the risks of ad libraries should not cause us to overlook their contribution to the Android ecosystem~\cite{leontiadis2012don}.
Unfortunately, there is currently very little in the way of norms or standards to protect this information.  While the Google Play Developer Distribution Agreement does allow Google to remove applications that are ``deemed to be ... spyware,'' the only mentions of privacy regard Google's ability to collect data on the developers, not on the developers' respect for user privacy~\cite{playDDA}.  It is clear that standards regarding user privacy are needed, whether they come from entities that manage application stores (such as Google) or from some other source.

\section{Conclusion}

API calls that allow the developer to expose personal information are present in most of the top 20 ad libraries.  They are generally designed to allow application developers to transmit demographic or targeting information that might be used to target ads at a given user.  While few libraries include API calls that would be sufficient, by themselves, to de-anonymize application users, in several occasions, the data provided by some combination of these calls could be sufficient to correlate the user with a real world identity.

Most mobile applications do not make use of these privacy related API calls.  However, the number which do choose to include these calls is not negligible.  As such calls are more common in more popular applications, and as ad agencies have the ability to correlate data from multiple applications, a significant portion of users have some personal data exposed through these API calls.

Unfortunately, users have no way to know of, approve, or block any transfer of information from applications to ad libraries and agencies.  Because application developers have an incentive to maximize ad revenue, they have a corresponding incentive to leak private user data, with few likely consequences.  If users are to have an expectation of privacy for data accessible from their mobile devices, some mechanism to report and manage these data flows is needed.
\newpage
\bibliographystyle{abbrv}
\bibliography{adApis}
\end{document}